\documentclass[aps,prd,11pt,notitlepage,nofootinbib,superscriptaddress,showkeys,letterpaper]{revtex4-1}
\usepackage{amstext,amsmath,amssymb,amsfonts,bbm, amsthm}
\usepackage[latin1]{inputenc}
\usepackage{fancyhdr}
\usepackage{graphicx}
\usepackage{hyperref}

\usepackage{epstopdf}
\usepackage{rotating}
\usepackage{cleveref}

\usepackage{float}

\usepackage{cases}
\usepackage{tabls}

\usepackage{subfigure}

\usepackage{amsthm}

\newtheorem{theorem}{Theorem}
\newtheorem{corollary}{Corollary}

\usepackage{tocvsec2}

\usepackage{color}

\def\beq{\begin{equation}}
\def\be{\begin{equation}}
\def\ee{\end{equation}}
\def\bes{\begin{eqnarray}}
\def\ees{\end{eqnarray}}

\begin{document}

\title{\large \bf Limiting fitness distributions in evolutionary dynamics}

\author{Matteo Smerlak}\email{matteo.smerlak@gmail.com}
\affiliation{Perimeter Institute for Theoretical Physics, 31 Caroline St.~N., Waterloo ON N2L 2Y5, Canada}
\author{Ahmed Youssef}\email{youssef@ld-research.com}
\affiliation{LD - Research, Pappelallee 78/79, 10437 Berlin, Germany}
\date{\small\today}

\begin{abstract}
Darwinian evolution can be modeled in general terms as a flow in the space of fitness (i.e. reproductive rate) distributions. In the limit where mutations are infinitely frequent and have infinitely small fitness effects (the ``diffusion approximation"), Tsimring \textit{et al.} have showed that this flow admits ``fitness wave" solutions: Gaussian-shape fitness distributions moving towards higher fitness values at constant speed. Here we show more generally that evolving fitness distributions are attracted to a one-parameter family of distributions with a fixed parabolic relationship between skewness and kurtosis. Unlike fitness waves, this statistical pattern encompasses both positive and negative (a.k.a. purifying) selection and is not restricted to rapidly adapting populations. Moreover we find that the mean fitness of a population under the selection of pre-existing variation is a power-law function of time, as observed in microbiological evolution experiments but at variance with fitness wave theory. At the conceptual level, our results can be viewed as the resolution of the ``dynamic insufficiency" of Fisher's fundamental theorem of natural selection. Our predictions are in good agreement with numerical simulations.
 \end{abstract}









\maketitle

\begin{quote}
	``In general, inference in biology depends critically on understanding the nature of limiting distributions." --- S. A. Frank \cite{Frank:2009ix}.
\end{quote}

\section{Introduction}
Evolution is the self-sustaining, open-ended process arising wherever entities (organisms, algorithms, memes, etc.) are subject to differential reproduction, heredity and variation. The specifics of this process are as diverse as life itself, but its basic structure---the principle of the ``survival of the fittest"---is abstract and universal. Does this general \textit{principle} translate into a general \textit{pattern}---a pattern which could be predicted mathematically and tested empirically, in the same way in which, say, energy conservation translates into the Boltzmann distribution in statistical mechanics? Are there falsifiable \textit{laws of evolution}?

Experience in the physical and statistical sciences suggests that emergent patterns often arise through suitable ``coarse-gainings", i.e. after large numbers of different configurations of the system are grouped according to dynamically relevant \textit{macroscopic} variables. In practice, pinpointing these relevant variables is no easy task, as the history of thermodynamics and the discovery of energy as a conserved quantity shows; in biology, it was not until Darwin's \textit{Origin of Species} that such a variable---reproductive rate or Malthusian \textit{fitness}\footnote{Malthusian fitness is also known as log-fitness, because it is the logarithm of the Wrightian fitness---the expected number of viable offsprings per capita. In this paper ``fitness" always refers to Malthusian fitness; the conversion to Wrightian fitness is immediate (normal distributions become log-normal etc.)}---was identified. The formal analogy between these two coarse-grained variables, energy and fitness, has been repeatedly highlighted in recent years \cite{PrugelBennett:1994uy,Sella2005}, but an equivalent to the Boltzmann distribution for evolutionary dynamics has yet to be identified.

These considerations encourage us to push Darwin's logic to its end and treat populations as collections of \textit{fitness classes}---groups of individuals with the same reproductive rate irrespective of their 	phenotype; we can then ask about the structure of \textit{fitness distributions} in evolving systems. To be sure, this approach amounts to a dramatic reduction of biological (or algorithmic, or cultural) reality: important ingredients such as genotype, phenotype, but also evolutionarily stable strategies, etc., are entirely left out of the analysis. Such may be, however, the price to pay to lift ourselves off from system-dependent properties and extract a robust prediction from the Darwinian paradigm. 


To our knowledge, the fitness space approach was first explored by Eshel in 1971 within a discrete-time framework \cite{Eshel:1971ur}; in continuous time, it was introduced by Tsimring \textit{et al.} \cite{Tsimring:1996cr}. Using a  ``diffusion approximation" familiar from non-equilibrium statistical mechanics, these authors went on to identify a class of ``fitness waves" solutions: Gaussian-shape fitness distributions moving at constant speed towards higher fitness values  \cite{Tsimring:1996cr}. Subsequent literature has developed methods to compute the speed of these fitness waves in terms of the frequency, effect and fixation probability of new mutations \cite{Rouzine:2003en,Desai:2007fh,Park:2010dz,Hallatschek:2011cu,Good:2012hm,Szendro:2013co}. The major finding of these studies is the extreme sensitivity of that speed to stochastic effects of rare mutations: in Daniel Fisher's words, fitness waves describe the evolutionary process as ``a dog led by its mutational nose" \cite{Fisher:2011fh}.
 
Fitness waves hint at a general statistical pattern, but they are incomplete in one key respect: they disregard the importance of \textit{negative selection} in evolution. Positive (or directional) selection consists in the growth of newly discovered high-fitness traits at the expense of slower reproducers, resulting in significant gains for the population mean fitness; by contrast, negative selection is the removal of low-fitness individuals without any new beneficial traits being introduced into the population. A condition for positive selection to sustain itself through time is that beneficial mutations are sufficiently frequent: the ``diffusion approximation" of Tsimring \textit{et al.} expresses this assumption in extreme form by imposing that new mutations are \textit{infinitely frequent} with \textit{infinitely small} fitness effects (both beneficial and deleterious). This can potentially capture aspects of the evolutionary process in phases of rapid adaptation, but not much more---real mutations (in biology and in other fields) are rare, mostly deleterious, and can occasionally have major fitness effects.

 In this paper we reconsider the dynamics of fitness distributions under less restrictive assumptions, with the goal of identifying more general patterns of evolution. Our approach consists in proving \textit{limit theorems} for fitness distributions, analogous to the H-theorem in statistical physics or the central limit theorem in probability theory. Limit theorems are powerful tools which cut through the complexity of statistical phenomena to extract their emergent, or ``universal" properties; in this sense they are the mathematician's ``extra sense" famously envied by Darwin.\footnote{``I have deeply regretted that I did not proceed far enough at least to understand something of the great leading principles of mathematics; for men thus endowed seem to have an extra sense." --- C. Darwin \cite{Darwin:1887vk}} A key ingredient underlying all limit theorems is the distinction between a probability distribution and its \textit{type}. Technically, the type of distribution is its equivalence class under an affine transformation of its argument; in practice, this means the \textit{shape} of the distribution, i.e. all information in the distribution except its location and scale. As a rule, limit theorems show that distribution types are more strongly constrained by large-number effects than location or scale. In the H-theorem, for instance, the mean and variance of the Boltzmann distribution depend on temperature, hence are system-dependent; its exponential structure is not.\footnote{The same is true in the central limit theorem: the mean and variance of the sum of many independent random variables depend of the variables' distributions, but their normal type does not.} That is, \textit{types, not distributions, are subject to emergence and universality}.
  
 We combine an exact solution of the general replication-mutation equation in fitness space with suitable asymptotic estimates to prove that \textit{evolving fitness distributions are attracted to a one-parameter family of universal types}. This pattern breaks down into two sub-patterns: under positive selection (of pre-existing variation or of new mutations), fitness is normally distributed; under negative selection and weak mutations, fitness has a reverted gamma distribution. A generic evolution trajectory consists of crossovers between these types. We check these findings with numerical simulations of the Wright-Fisher process and with a simple genetic algorithm.

\section{Evolution as transport}

We consider an infinite population with a continuous  distribution of (Malthusian) fitness $p_t(x)$. We assume that fitness-changing new mutations occur with a rate $U$, and that their fitness effects $x\mapsto x+\Delta$, with $\Delta>0$ (resp. $\Delta<0$) for beneficial (resp. deleterious) mutations, are distributed according to some distribution of fitness effects (DFE) $m(\Delta)$, a common assumption in evolutionary biology \cite{EyreWalker:2007dl}. (For clarity of the presentation we assume that the mutation rate $U$ and DFE $m(\Delta)$ are fixed, but our results can be straightforwardly generalized to time-dependent and/or stochastic mutational effects to allow for changing fitness landscapes, changing environments, etc.)

The dynamical equation for the evolution of fitness distributions under selection and mutation can be formulated both in continuous and in discrete time, without it making any difference for our purposes. For definiteness we focus on the continuous case,\footnote{See the Appendix for the corresponding results in discrete time.} where it reads 
\begin{equation}\label{floweq}
	\frac{\partial p_t(x)}{\partial t}=(x-\mu_t)p_t(x)+U\int d\Delta\, m(\Delta)\,[p_t(x-\Delta)-p_t(x)]
\end{equation}
with $\mu_t\equiv\int dx\, x\,p_t(x)$ the mean fitness at time $t$. The first term in this equation expresses natural selection, i.e. the population effect of differential fitness; the second term is the mutation term, responsible for the introduction of new variations in fitness distributed according to the DFE $m(\Delta)$.   

\begin{figure}
	\includegraphics{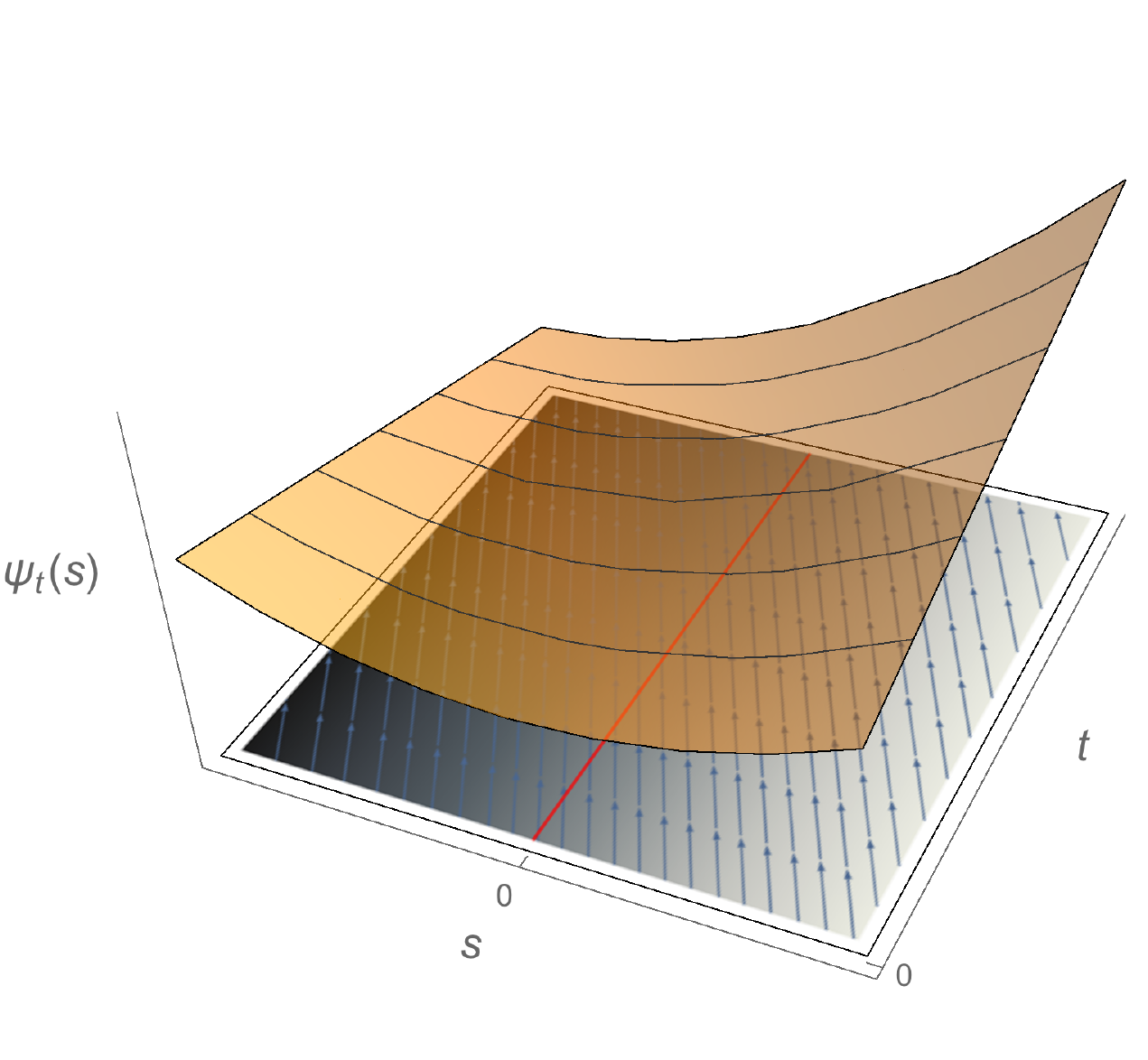}
	\caption{Evolution as transport in $(t,s)$ space. The CGF $\psi_t(s)$ (curved surface) is transported along the characteristics $s+t=\textrm{constant}$ (arrows), with a mutational source term $U(\chi_m(s)-1)$ (grey shading, with darker colors indicating larger values of $\chi_m(s)$, hence stronger mutational effects). The cumulants of the fitness distribution are given by the $s$-derivatives of $\psi_t(s)$ at $s=0$ (red line). For this plot we took $p_0(x)=S_{0,1,0}(x)$ and $m(\Delta)=S_{0,.1,-1}(\Delta)$ with $S_{\mu,\sigma,\xi}$ the skew-normal distribution with location $\mu$, scale $\sigma$ and shape $\xi$.}
	\label{transportfig}
\end{figure}

After Ref. \cite{Tsimring:1996cr} this ``replicator-mutator equation"\footnote{Replicator-mutator equations are often written in genotype or trait space \cite{Kimura:1958gd,Eigen:1971cf}; here by contrast it is introduced as an equation for fitness distributions.} is often approximated by a reaction-diffusion equation, with the mutation integral replaced by a term proportional to $\partial^2p_t(x)/\partial x^2$, but this step is neither well justified---real mutations are not infinitely frequent---nor necessary. Indeed we can obtain the general solution of \eqref{floweq} in closed form for any DFE. It suffices for that to transform \eqref{floweq} into an equation for the cumulant-generating-function (CGF) of the fitness distribution, defined by $\psi_t(s)\equiv\ln[\int dx\, e^{sx}p_t(x)]$. This gives a simple transport-like equation, with explicit solution (Appendix \ref{appendixevo})
\begin{equation}\label{solution}
	\psi_t(s)=\psi_0(s+t)-\psi_0(t)+U\int_0^tdu\, \big(\chi_{m}(s+u)-\chi_{m}(u)\big)
\end{equation}
where $\chi_m(s)\equiv\int d\Delta\, e^{s\Delta}m(\Delta)$ is the moment-generating-function (MGF) of the DFE. This reformulation is intuitively appealing: in $(t,s)$-space, natural selection corresponds to the transport of the initial CGF $\psi_0(s)$ towards progressively lower values of $s$, and mutations to a source term, see Fig. \ref{transportfig}. 

The explicit solution \eqref{solution} allows us to distinguish between two different regimes of evolution. Indeed depending on the relative magnitude of the terms above, the evolutionary trajectory can either be dominated by the \textit{selection of pre-existing variation}, in which case
\begin{equation}\label{preexisting}
	\psi_t(s)\simeq\psi_0(s+t)-\psi_0(t),
\end{equation}
or by the \textit{selection of new mutations}, and then we have
\begin{equation}\label{new}
	\psi_t(s)\simeq U\int_0^tdu\, \big(\chi_m(s+u)-\chi_m(u)\big).
\end{equation}
Note that Eq. \eqref{new} is \textit{not} the solution of \eqref{floweq} without the selection term on the right-hand side.


\section{Four limit theorems}\label{limittheorems}
We now analyze these two regimes separately, using the tools of asymptotic analysis. 
Henceforth the symbol $\sim$ stands for ``asymptotically equivalent up to a multiplicative constant".

\subsection{Selection of pre-existing variation}\label{preexistingsel}
We begin by focusing on the case where Eq. \eqref{preexisting} holds, either because the mutation rate $U$ is small or because the initial fitness distribution $p_0(x)$ is broad. This regime, the selection of pre-existing variation, is covered by the two limit theorems below.\footnote{In the final stages of this work we discovered that some of the results in Sec. \ref{preexistingsel} are contained in theorems obtained in a very different context by Balkema, Kl\"uppelberg, and Resnick \cite{Balkema:1999we,Balkema:2003hn}.}

The first theorem is concerned with initial fitness distributions with unbounded support, i.e. such that $\sup_x\{p_0(x)>0\}=\infty$. This condition is the mathematical counterpart of the biological notion of \textit{positive} (or directional) selection, expressing the idea that ever-higher fitness individuals continuously take over the population at the expense of more common but less fit variants.

\begin{theorem}[Positive selection of pre-existing variation]
	
If Eq. \eqref{preexisting} holds and the initial fitness distribution $p_0(x)$ has unbounded support with $$-\ln\int_x^\infty dy\,p_0(y)\underset{x\to\infty}{\sim} x^\alpha$$ for any $\alpha>1$, then the fitness distribution $p_t(x)$ becomes asymptotically normal as $t$ grows. Furthermore, the mean $\mu_t$ and variance $\sigma_t^2$ of $p_t(x)$ scale as  
$$	\mu_t\underset{t\to\infty}{\sim}  t^{\overline{\alpha}-1}\quad\textrm{and}\quad\sigma^2_t\underset{t\to\infty}{\sim} t^{\overline{\alpha}-2}$$ where $\overline{\alpha}$ is the exponent conjugate to $\alpha$, i.e. $1/\alpha+1/\overline{\alpha}=1$.
\end{theorem}
We make several remarks concerning this limit theorem. First, and in spite of an obvious formal similarity, Thm. 1 is \textit{not} a consequence of the central limit theorem. This is apparent from its proof in Appendix \ref{proofs}, which involves a different kind of asymptotic estimate; it is also clear from the scaling behavior of the mean $\mu_t$ and variance $\sigma_t^2$: in contrast with the central limit theorem, where the mean and variance are linear functions of the sample size, here $\mu_t$ and $\sigma_t^2$ do \textit{not} grow with time at the same rate. Second, the thin-tail condition on $p_0(x)$ in Thm. 1 is a general one, which encompasses many classical distributions (such as the Weibull family); it can moreover be generalized further in terms of the notion of  ``regular variation at infinity" \cite{BINGHAM:1989ca}. Third, a lower bound on the rate of convergence of the fitness distribution to the normal type can be estimated as a function of the tail index $\alpha$, see Appendix \ref{proofs}. 

Our second theorem deals with the case of fitness distributions with a finite right endpoint $x_+\equiv\sup_x\{p_0(x)>0\}<\infty$, corresponding to the \textit{negative} selection of pre-existing variation.

\begin{figure}
\begin{minipage}{0.45\linewidth}
			\includegraphics[scale=.45]{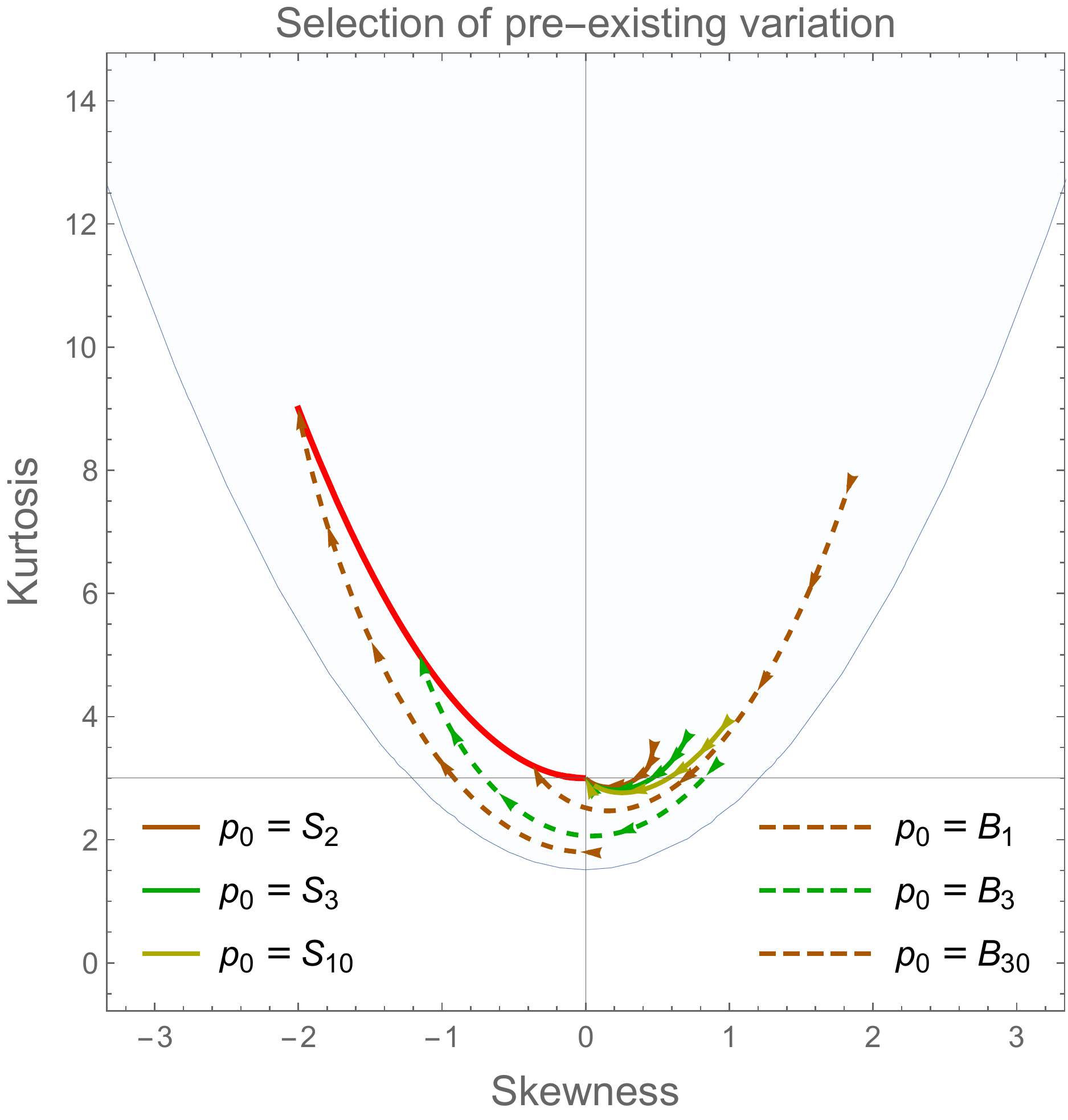}	
\end{minipage}
\hfill
\begin{minipage}{0.45\linewidth}
			\vspace{0.2cm}
			\includegraphics[scale=.33]{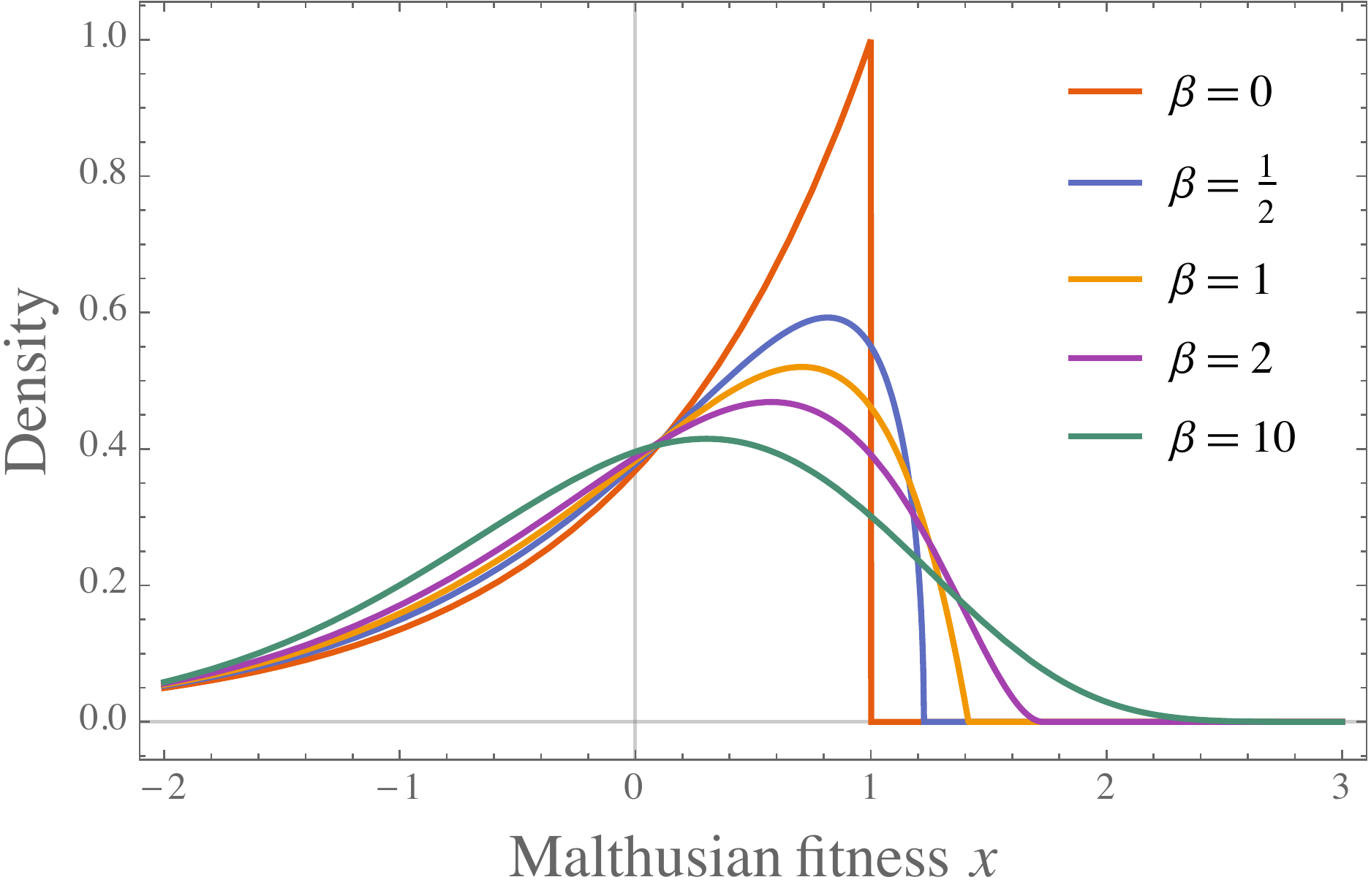}\vfill
		\includegraphics[scale=.33]{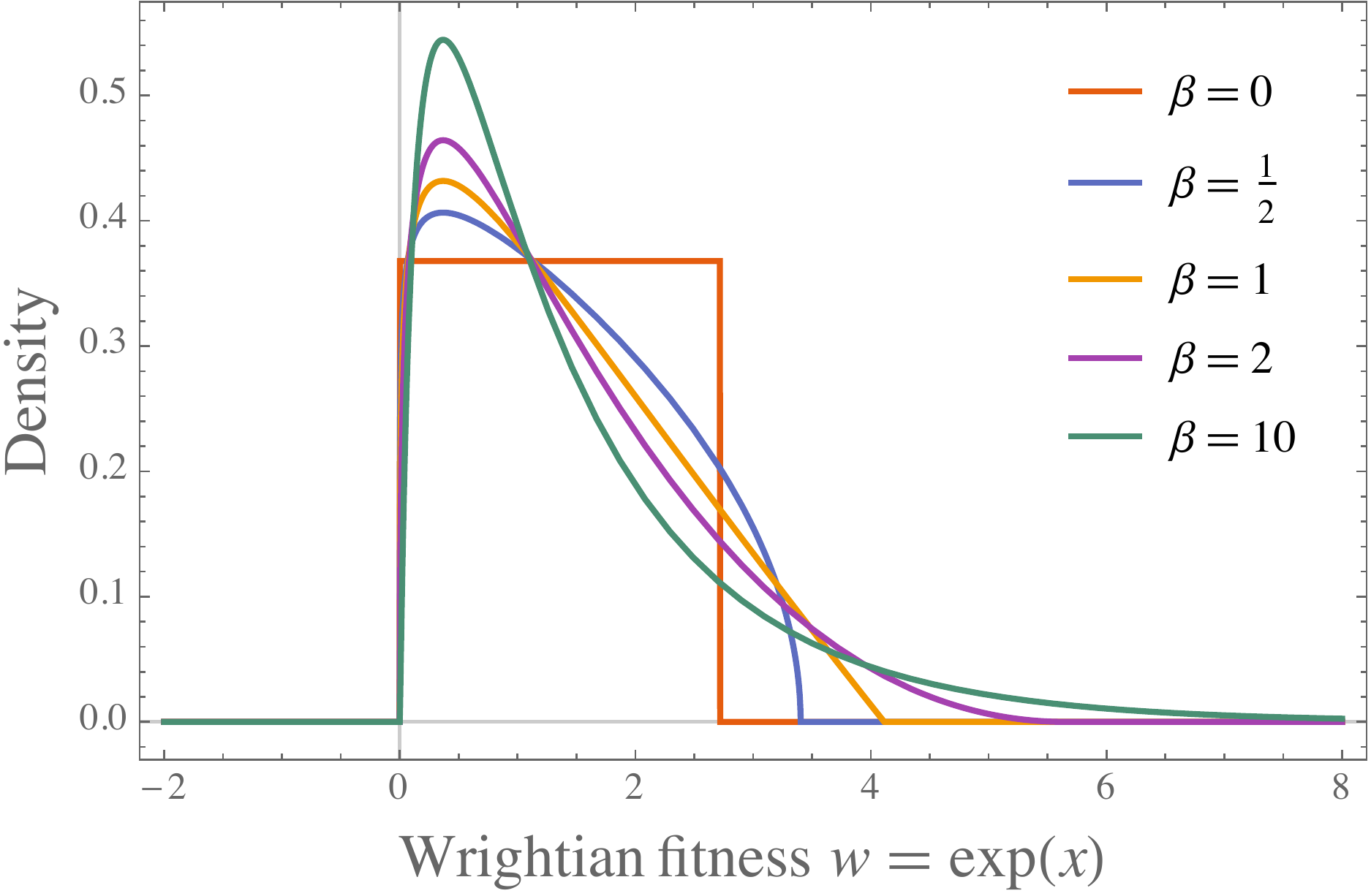}
\end{minipage}
	\caption{Attractors for the selection of pre-existing variation, described by Eq. \eqref{preexisting}. Left: convergence of various initial distributions to the $g_\beta$ family (red curve) in the skewness-kurtosis plane. The continuous lines correspond to initial distributions with unbounded support (three skew-normal distributions $S_\xi$ with shape parameter $\xi$); the dashed lines correspond to initial distributions with bounded support (three beta distributions $B_\xi$ with shape parameter $\xi$.) Right: shape of the attracting distributions $g_\beta$ for various values of $\beta$, in terms of Malthusian fitness $x$ (top) and of Wrigtian fitness $w=e^x$ (bottom).
}
	\label{plotnewdist}
\end{figure}

\begin{theorem}[Negative selection of pre-existing variation]
	
If Eq. \eqref{preexisting} holds and $p_0(x)$ has a finite right endpoint $x_+$ with $$p_0(x_+-x)\underset{x\to x_+}{\sim} (x_+-x)^\beta$$ for some $\beta\geq 0$, then the fitness distribution $p_t(x)$ becomes asymptotically a reversed Gamma distribution with shape parameter $1+\beta$ as $t$ grows, i.e. converges in type to the distribution with density function
		$$g_\beta(x)\equiv\frac{(1+\beta)^{(1+\beta)/2}}{\Gamma(1+\beta)}\,e^{-(1+\beta)^{1/2}[(1+\beta)^{1/2}-x]}\Big[(1+\beta)^{1/2}-x\Big]^\beta \quad\textrm{for}\quad x\leq (1+\beta)^{1/2}.
$$
Furthermore, the variance in fitness $\sigma_t^2$ eventually decreases as $\sigma_t^2\sim t^{-2}.$
\end{theorem}
The reversed gamma distributions above interpolate between a reversed exponential distribution for $\beta=0$ and a normal distribution for $\beta\to\infty$. In this sense, the situations of positive selection (Thm. 1) and negative selection (Thm. 2) are unified in a single continuous one-parameter family of distribution types, plotted in Fig. \ref{plotnewdist}. Given this, we can interpret the parameter $1/\beta$ as \textit{selection negativity}: low values of $\beta$ correspond to highly skewed distributions strongly dominated by the high-fitness individuals; high values of $\beta$, on the other hand, correspond to the situation where the most frequent individuals in the population have sub-optimal fitness.  
  
%
%


\subsection{Selection of new mutations}

In the second regime of evolution, captured mathematically by Eq. \eqref{new}, the structure of the fitness distribution is determined by the constant stream of new mutations rather than by initial conditions. The nature of evolutionary dynamics in this regime depends on whether these mutations are sometimes beneficial (positive selection) or always deleterious or neutral (negative selection). 


Our next theorem shows that, provided at least some mutations are beneficial, $\Delta_+\equiv \sup_\Delta \{m(\Delta)>0\}>0$, the fitness distribution converges to the normal type \textit{for any DFE}. This is a strong form of universality in evolutionary dynamics. 

\begin{theorem}[Positive selection of new mutations]
If Eq. \eqref{new} holds and at least some mutations are beneficial, $\Delta_+> 0$, the fitness distributions becomes asymptotically normal as $t$ grows independently of the distribution of fitness effects. 
\end{theorem}

Finally, the case where all mutations are deleterious or neutral ($\Delta_+\leq 0$) was treated by Eshel in the context of discrete generations \cite{Eshel:1971ur}. We obtain the following result.

\begin{theorem}[Mutation-selection balance]
If Eq. \eqref{new} holds and all mutations are either deleterious or neutral, $\Delta_+\leq 0$, the fitness distributions $p_t(x)$ converges to the unique distribution with mean $\mu_\infty=x_+-U$, variance $\sigma_\infty^2=U\mu_m$ and higher standardized cumulants $$K_\infty^{(p)}=-U^{1-p/2}\,\frac{\mu_m^{(p-1)}}{\mu_m^{p/2}}\quad\textrm{for}\quad p\geq 3$$ 
where $\mu_m$ is the mean fitness effect and $\mu_m^{(p)}$ the higher moments of the DFE $m(\Delta)$. In particular:
\begin{itemize}
	\item  the asymptotic mean fitness is independent of the distribution of fitness effects, as noted by Eshel \cite{Eshel:1971ur}, and
	\item the asymptotic fitness distribution becomes normal in the limit of large mutation rates ($U\to\infty$) or small fitness effects ($\sigma_m\to 0$). 

\end{itemize}
\end{theorem}

Note that, unlike the situation in Theorems 1-3, which all describe to phases of \textit{adaptative evolution} (the mean fitness increases), the mutation-selection balance in Theorem 4 does \textit{not} lead to universal fitness distributions, except when fitness effects are very small or mutation rates very high.

\section{Discussion}

\subsection{Crossovers}

The theorems above capture the dynamics of fitness distributions in different limits, none of which holds exactly true in a real system. However these idealizations can serve as a basis to describe realistic evolutionary trajectories. Consider a large population starting with some pre-existing variation in fitness $\sigma_0$, subject to rare mutations (small $U$) with small fitness effects (small $\sigma_m$). Moreover suppose that the initial mean fitness $\mu_0$ is far from the maximal fitness $x_+$ available to the system, $\mu_0\ll x_+$. The qualitative behavior of the fitness distribution $p_t(x)$ at future times is then completely prescribed by Theorems 1-4. 

In a first phase of evolution, the dynamics is dominated by the positive selection of pre-existing variation and the fitness distribution converges to the normal type, with the scaling of its mean $\mu_t$ and variance $\sigma_t^2$ determined by the high-fitness tail of $p_0(x)$, as stated in Thm. 1. When the of pre-existing variation is exhausted, i.e. when $x_t\simeq x_+$, selection becomes negative: the type of the distribution undergoes a rapid \textit{crossover} to one of the $g_\beta(x)$ distributions in Thm. 2; the variance in fitness then starting to decay as $\sigma_t^2\sim t^{-2}$. This situation continues until mutations start becoming dominant, $\sigma_t\simeq \mu_m$. At that point, the future behavior of the fitness distribution depends on the whether mutations are partly beneficial ($\Delta_+>0$) or entirely deleterious ($\Delta_+\leq 0$). In the former case, Thm. 3 applies and the distribution undergoes a second crossover,  taking it back to the normal type; in the latter case, a non-universal mutation-selection balance is reached, with $p_t(x)$ converging to a fixed limit $p_\infty(x)$ entirely determined by the DFE $m(x)$. 

\begin{figure}
	\includegraphics[scale=.4]{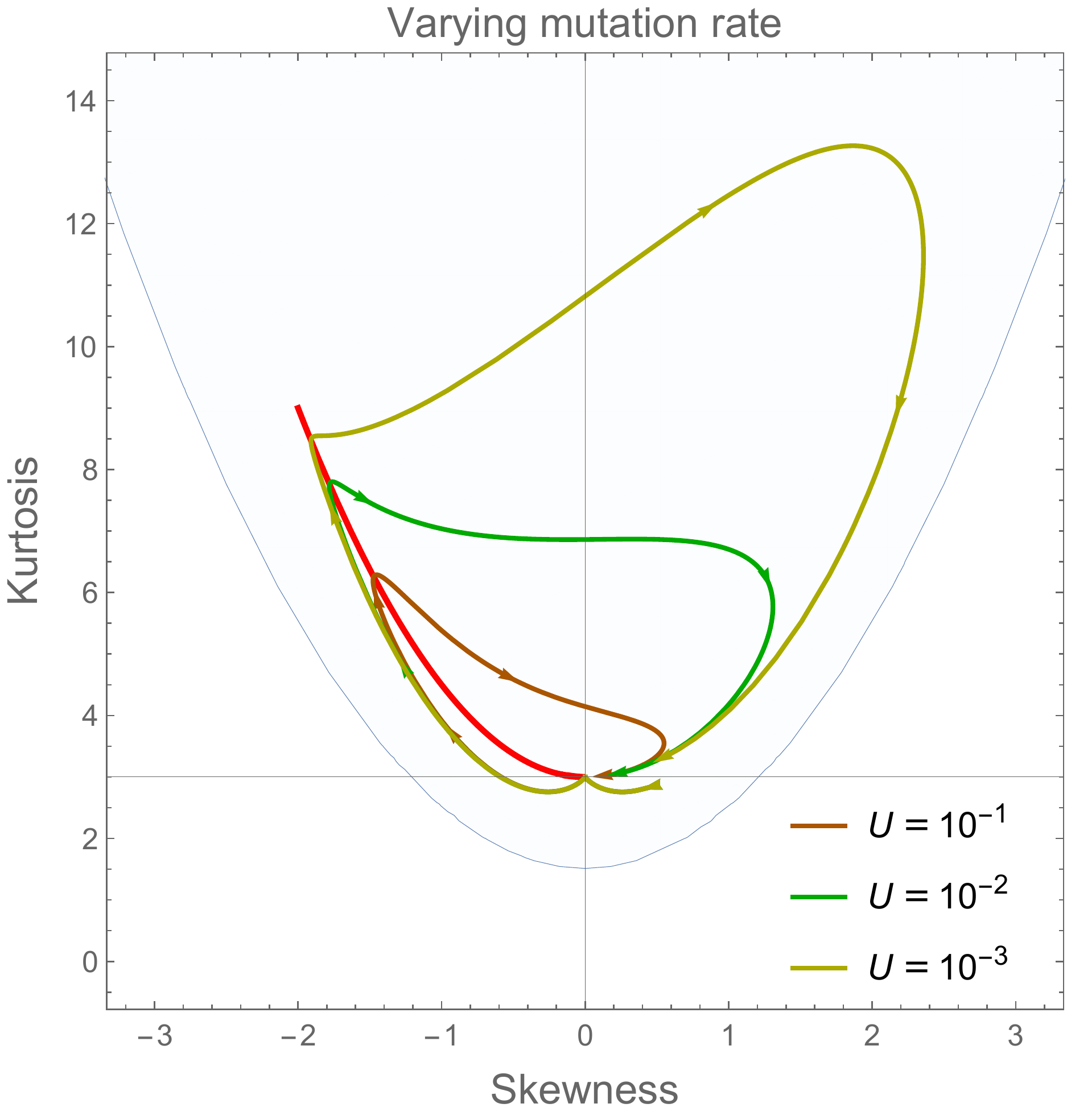}\hfill
	\includegraphics[scale=.4]{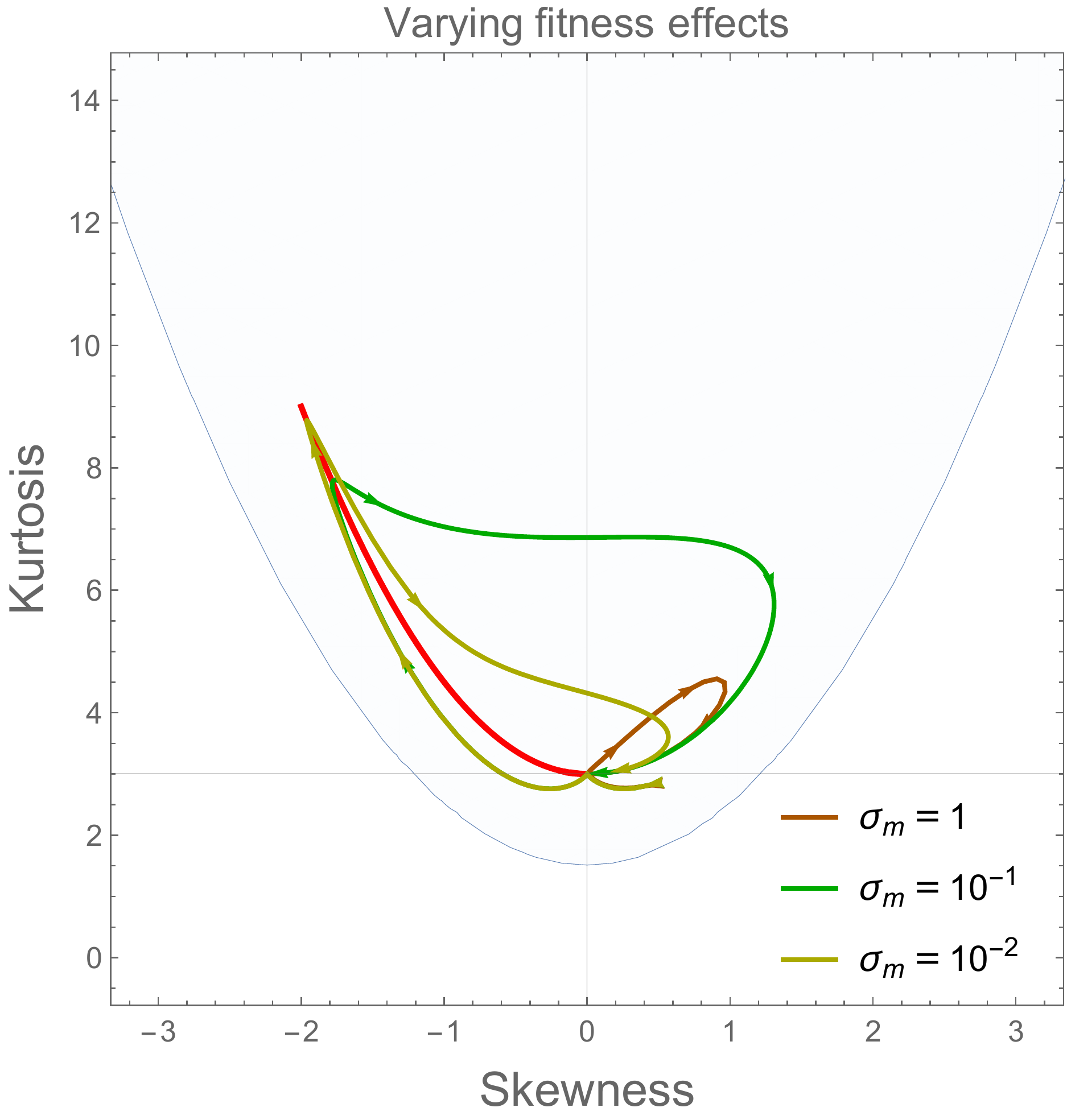}
	\caption{Crossover between different attractors, as follows from to the general solution \eqref{solution}	. Here we took a standard normal distribution truncated at $x_+=10$ as initial fitness distribution and as skew-normal distribution with location $0$, scale $\sigma_m$ and shape parameter $-1$ as DFE. The blue-shaded region is the subset of the skewness-kurtosis plane available to unimodal distributions according to the Klaassen-Mokveld-van Es inequality \cite{Klaassen:2000ur}. The red curve is the $\beta$-family of attractors in Thm. 1-3, with then normal at $(S,K)=(0,3)$ corresponding to positive selection of pre-existing variation and of new mutations. Note that the attractiveness of the red curve is an increasing function of $U/\sigma_m$.}
	\label{plotcrossoverSK}
\end{figure}

This sequence of patterns and crossovers between them is illustrated in the skewness-kurtosis plane in Fig. \ref{plotcrossoverSK}. The skewness $S_t$ and kurtosis $K_t$ of the fitness distributions are defined in terms of its third and fourth cumulants $\kappa_t^{(3)}$ and $\kappa_t^{(4)}$ as $S_t\equiv \kappa_t^{(3)}/\sigma_t^3$ and $K_t\equiv\kappa_t^{(4)}/\sigma_t^4$; this plane allows a very convenient low-dimensional representation of  distributions types.

\subsection{Dynamic sufficiency and Fisher's fundamental theorem}

It is well known (and easy to check) that the dynamical equation \eqref{floweq} is equivalent to the infinite tower of cumulant equations
\begin{equation}\label{tower}
	\frac{\partial\kappa_t^{(p)}}{\partial t}=\kappa_t^{(p+1)}+U\mu_m^{(p)} \quad\textrm{for}\ \ p\geq 1
\end{equation}
where $\kappa_t^{(p)}$ is the $p$-th cumulant of the fitness distribution and $\mu_m^{(p)}$ is the $p$-th moment of the DFE. The equation for $p=1$, which relates the growth of the mean fitness to the variance in fitness (and mutational effects), is known as \textit{Fisher's fundamental theorem of natural selection} \cite{Fisher:1930wy}. The tower of equations \eqref{tower} being unclosed, it has been claimed that Fisher's theorem (or its generalization) is ``dynamically insufficient", i.e. that it has no predictive power; see \cite{Frank:2012dz} and references therein for a review of this literature.

Our results in this paper show that this view is overly pessimistic. To begin with, the solution \eqref{solution} translates into exact expressions for the cumulants in terms of the initial CGF and of the DFE, namely (Appendix \ref{appendixevo})
\begin{equation}
	\kappa_t^{(p)}=\psi_0^{(p)}(t)+U\big(\chi_m^{(p-1)}(t)-\chi_m^{(p-1)}(0)\big)\quad\textrm{for}\ \ p\geq 1.
\end{equation}
While not solely expressed in terms of cumulants, this solution does predict the future evolution of all $\kappa_t^{(p)}$ given an initial fitness distribution and a DFE. Secondly, Theorems 1-3 imply very tight asymptotic relationships between the cumulants of the fitness distribution along an adaptive evolutionary trajectory. In particular, 

\begin{corollary} In each one of the three adaptive regimes described by Theorems 1-3, the skewness $S_t\equiv\kappa_t^{(3)}/\sigma_t^3$ and kurtosis $K_t\equiv \kappa_t^{(4)}/\sigma_t^4$ of the fitness distribution are attracted towards the universal relationship  
	$K_t\sim3+\frac{3}{2}S_t^2$ with $-2\leq S_t\leq 0$.
\end{corollary}
This universal relationship---consistent with the mathematical inequality $K\geq S^2 +189/125$ holding for any unimodal distribution \cite{Klaassen:2000ur}---is a property of gamma and normal distributions. Along an evolutionary trajectory, it holds true at all times when the mean fitness increases (except during rapid crossover phases). Crucially, this relationship closes the tower of cumulant equations \eqref{tower}: the mean fitness is determined by the variance in fitness, which is determined by the skewness, which is determined by the kurtosis---which in turn is determined by the skewness.

\subsection{Scaling and drift}

We have emphasized that robust patterns should not be expected at the level of the mean fitness $\mu_t$ or the variance in fitness $\sigma_t^2$. This is intuitively clear: the mean fitness of a population depends on the contingent history of mutational effect along its particular evolutionary trajectory---a single strongly beneficial mutation can have major effects on the whole population. The recent literature on fitness waves cited in the introduction has also strongly emphasized this point. For this reason, we have not attempted in this paper to characterize the dynamics of $\mu_t$ and $\sigma_t^2$ in any detail: this would require a stochastic treatment dealing with the statistics of fixation and drift.

There is however one case is which we expect the infinite-population limit to be relevant for the evolution of the mean fitness and its variance: the selection of pre-existing variation. In this case, indeed, we can assume that all mutations have already been fixed in the population. In this regime, Thm. 1 implies that both $\mu_t$ and $\sigma_t^2$ should be \textit{power-law} functions of time, with an exponent which depends on the fat-tailedness of the initial fitness distribution (captured by the tail index $\alpha$). This prediction is at variance with fitness wave theory, which predicts a linear growth of the mean fitness \cite{Tsimring:1996cr}, but it matches with the results of a long-term evolution experiment with \textit{E. coli} \cite{Wiser:2013ch}. Whether pre-existing variation in fitness was indeed the main determinant of the fitness gains observed by Lenski \textit{et al.}, or new mutations did play a critical role in the evolution of \textit{E. coli} populations, is unclear to us. 

Also note that Thm. 1 implies the existence of a critical value of the tail index, $\alpha=2$, below which the variance in fitness \textit{increases} under pure selection. This means that, if the initial distribution of fitness is sufficiently fat-tailed, selection does not necessarily imply a loss of variation in fitness. It would be interesting to see if this somewhat counter-intuitive behavior is realized in real evolving systems, at least for sufficiently long transients during which new mutations do not play a significant role.

\section{Numerical experiments}

\begin{figure}
\begin{minipage}{0.35\linewidth}
			\includegraphics[scale=.33]{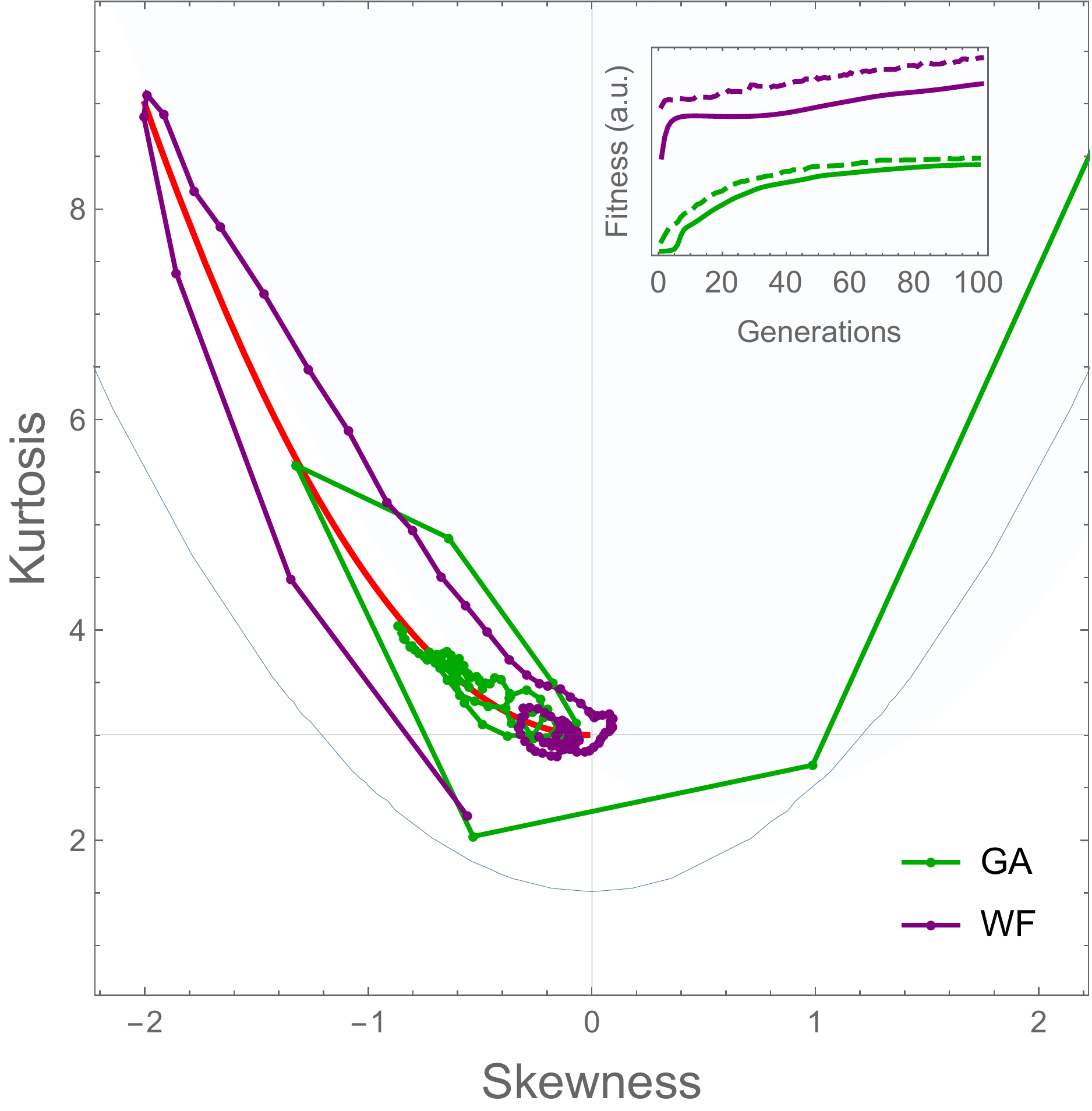}	
\end{minipage}
\hfill
\begin{minipage}{0.6\linewidth}
\vspace{-.3cm}
\includegraphics[scale=.25]{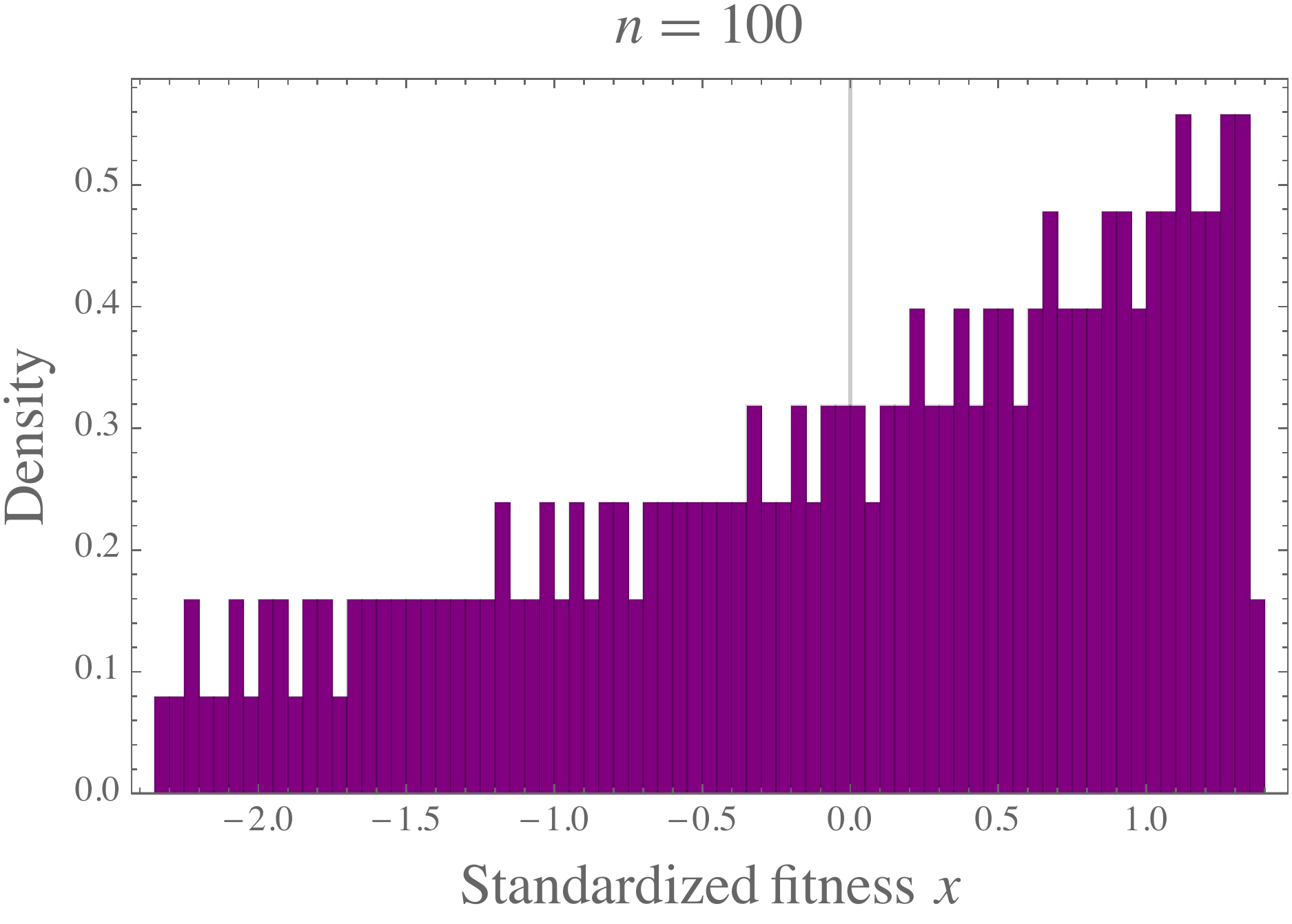}
\includegraphics[scale=.25]{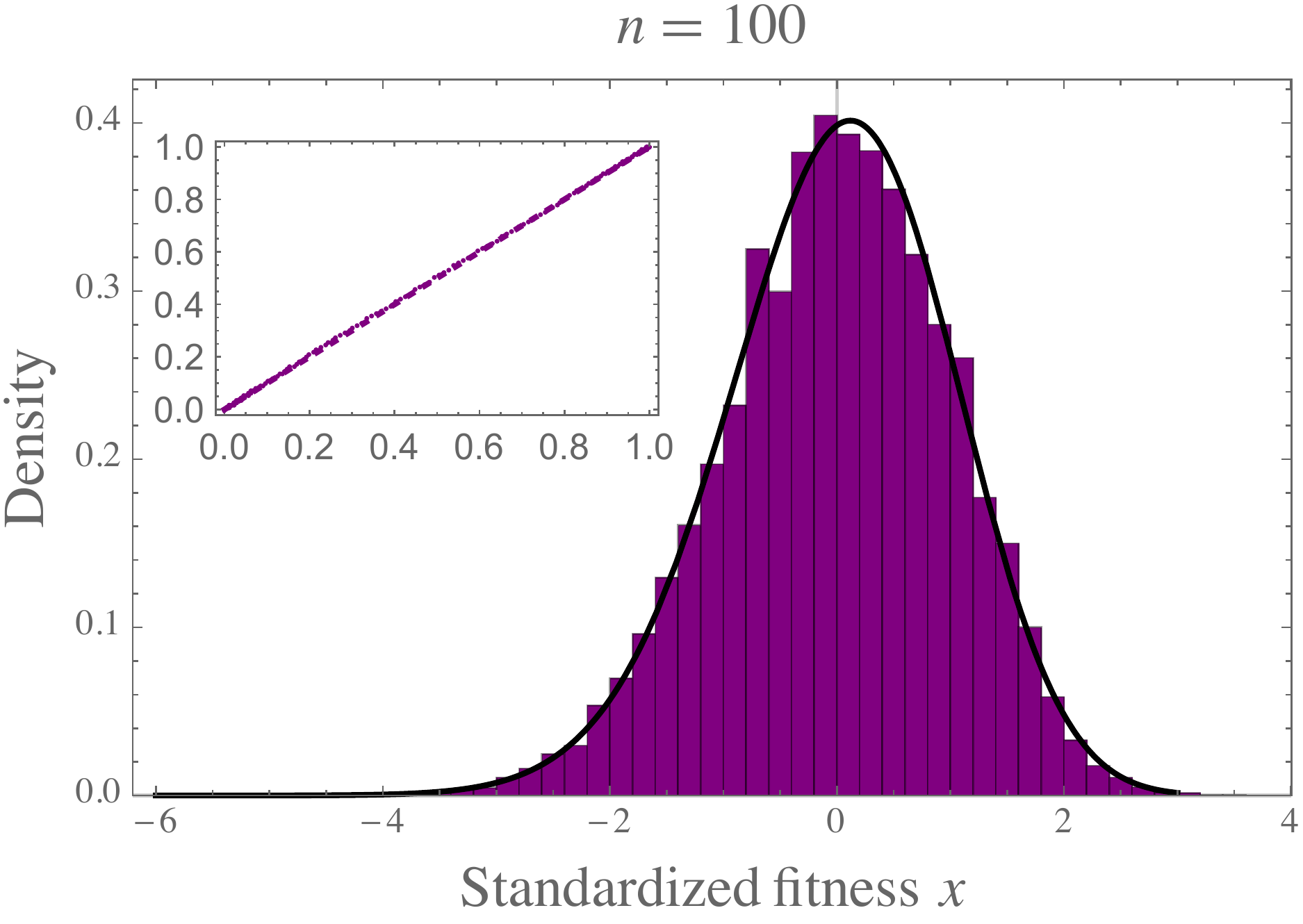}
\includegraphics[scale=.25]{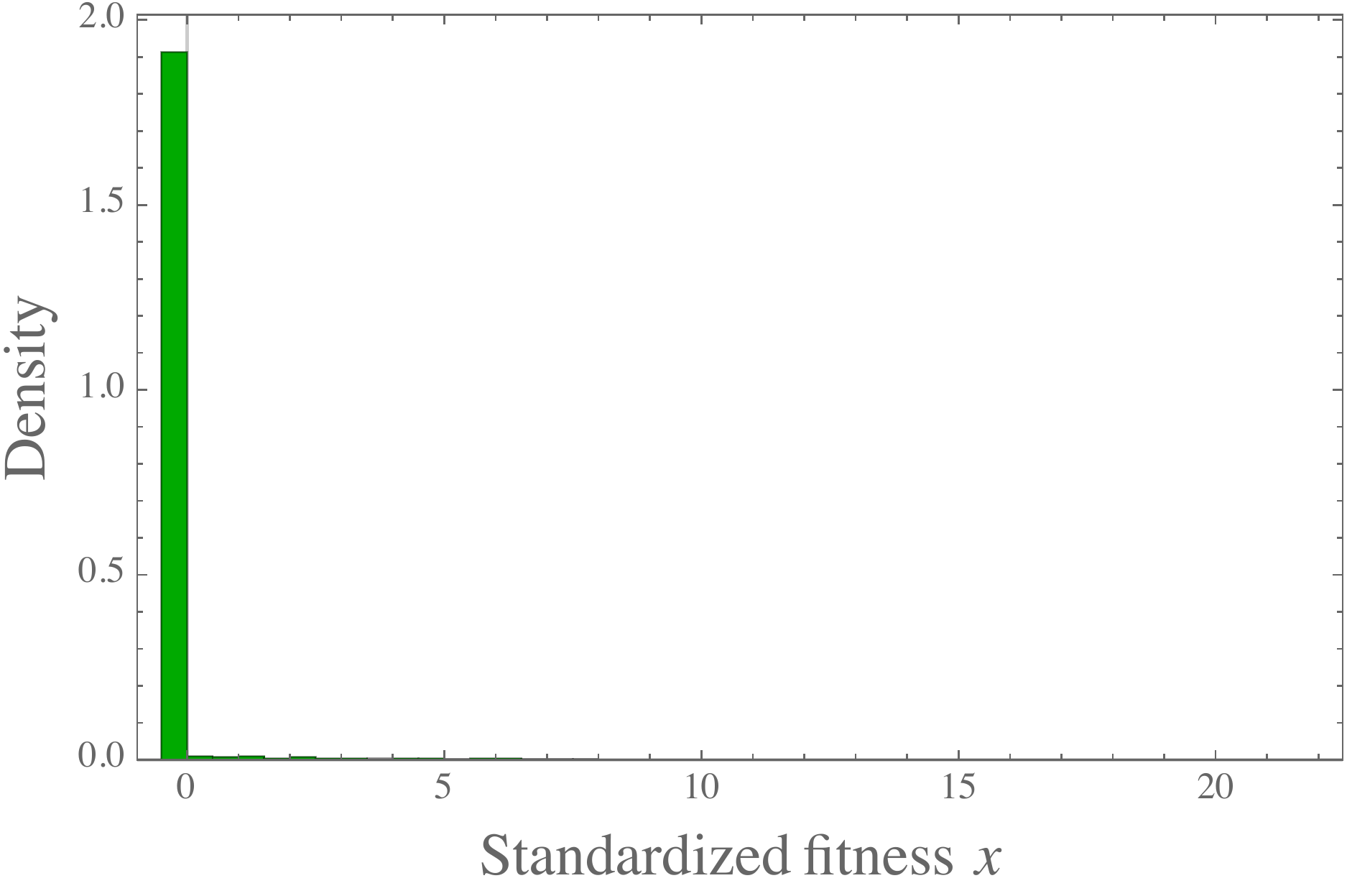}
\includegraphics[scale=.25]{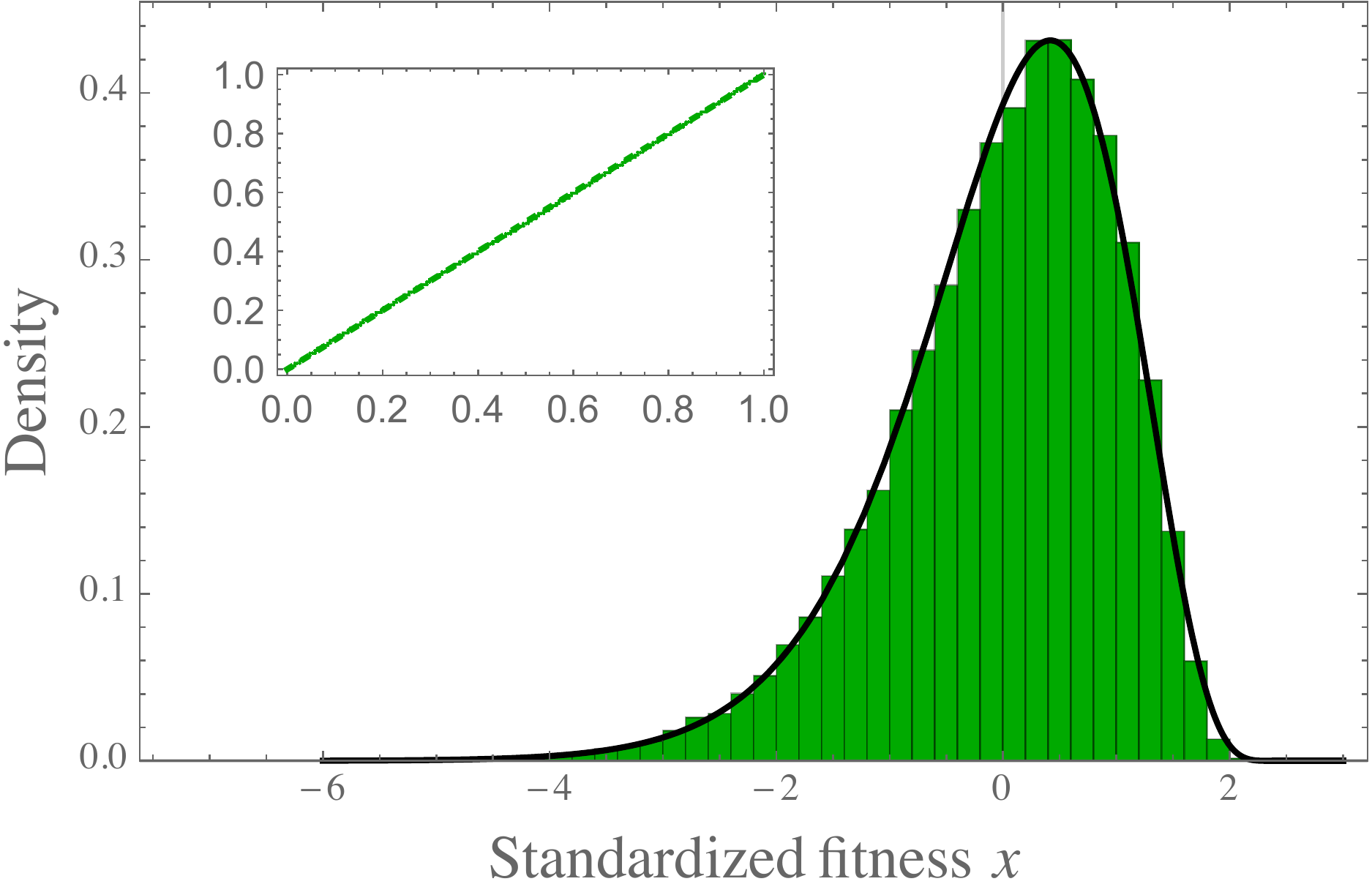}
\end{minipage}
	\caption{Numerical simulations with the WF process (purple) and with a simple GA (green). Left: trajectories of fitness distributions in the skewness-kurtosis plane for the WF and GA simulations described in the text, with the $g_\beta(x)$ attractors represented as the red parabolic curve; the inset plots the mean (continuous) and maximum (dashed) fitness in the population as a function of time. Right: the initial and final fitness distributions, overlaid with a best-fit distribution from the $g_\beta(x)$ family; the insets are P-P plot showing the good match between the empirical distributions and their fit.}
	\label{sims}
\end{figure}

\subsection{Wright-Fisher simulations}

We tested our results with Wright-Fisher (WF) simulations (Appendix \ref{WFappendix}). The WF process is a Markov chain representing the evolutionary dynamics of non-overlapping generations; it is a useful model to assess the importance of finite-population (drift) effects. We considered a (purposely relatively small) population of $10^5$ individuals grouped in $500$ distinct fitness classes, with Wrightian fitness (number of offspring) $w=e^x$ ranging between $w_{\textrm{min}}=1$ and $w_{\textrm{max}}=10$. Mutations were introduced with a rate $U=10^{-3}$, and their distribution of fitness effects $w\mapsto w'$ was assumed to be a fixed function of the selection coefficient $w'/w-1$, namely a Laplace distribution with location $-10^{-2}$ and scale $10^{-2}$. The fitness of the initial population was uniformly distributed between $w_{\textrm{min}}$ and the mid-point $(w_{\textrm{min}}+w_{\textrm{max}})/2$. We let this population evolve for $100$ generations and extracted the skewness and kurtosis of the resulting Malthusian fitness distributions.

The results, plotted in Fig. \ref{sims}, are in good agreement with the patterns predicted in sec. \ref{limittheorems}. The pre-existing variation in fitness was selected first, and the fitness distribution was correspondingly attracted to the reverted exponential, with skewness $S=-2$ and kurtosis $K=6$. After this initial adaptation phase, mutations started feeding new fitness gains, which translated into the fitness distribution being attracted to the normal type with $(S,K)=(0,3)$. A fit of the final standardized distribution, at generation $n=100$, with the parametric family $g_\beta(x)$ gave a best-fit value $\beta=69.6$ with a value of $9.10^{-3}$ for the Kolmogorov-Smirnov statistic.\footnote{The corresponding $p$-value is irrelevant here: finite-population errors are \textit{not} mere sampling errors from a fixed underlying distribution---the distribution itself is stochastic.}

\subsection{Genetic algorithm}
The relevance of our limit theorems is not restricted to biology. To illustrate this point we ran a genetic algorithm (GA) solving the linear integer optimization problem  (Appendix \ref{GAappendix})
\begin{equation}
	\max\{c\cdot y\,;\, y\in\{1,\cdots,Q\}^L\ \textrm{and}\ b\cdot y\leq d\}
\end{equation}
where $L,Q\in\mathbb{N}$ and $d>0$ are fixed numbers, $b$ and $c$ are two randomly chosen vectors in $[0,1]^L$, and $\cdot$ denotes the dot product of vectors. We coded $10^5$ candidate solutions to this problem (with $Q=10$, $L=100$ and $d=1$) as strings of integers (``bases"), and generated an evolutionary trajectory from a random initial population with a mutation rate per base per generation $u=10^{-2}$. Unlike the WF model, this is a setting where the evolutionary dynamics is not \textit{a priori} guaranteed to follow the replicator-mutator dynamics; in particular the DFE is unknown and possibly ill-defined (Appendix \ref{GAappendix}). In spite of this, we found that the fitness distribution quickly converged to the $g_\beta(x)$ attractors, see Fig. \ref{sims}. In this case, pre-existing variation in fitness was negligible and no crossover was observed. A fit of the empirical fitness distribution after $n=100$ generations with the $g_\beta(x)$ family gave $\beta=4.7$ with a value of $6.10^{-3}$ for the Kolmogorov-Smirnov statistic.

\section{Conclusion}

\begin {table}[b]
\begin{center}
\begin{tabular}{|c|c|c|}
 \hline
 \  selection \ & \ pre-existing variation \ & \ new mutations\ \\
  \hline
  positive &  normal & normal  \\
    negative & reverted gamma & \ DFE-dependent (but normal if $\sigma_m/U\to 0$)\ \\
           \hline
\end{tabular}
\caption {Four modes of natural selection and the types of the corresponding fitness distributions.}
\end{center}
\label{tabtab}
\end {table}

We have studied the dynamics of fitness distributions in four regimes of evolution, as summarized in Table I. In the regimes where evolution is adaptive, the attracting types belong to a single one-parameter family of distributions, parametrized by a ``selection negativity" parameter. In terms of Malthusian fitness, these distributions are all negatively skewed and leptokurtic, with a universal parabolic relationship between kurtosis and skewness. We argue that these limiting distributions play a role similar to that of the Boltzmann distribution in equilibrium statistical mechanics or the normal distribution in statistics: they act as a benchmark against which evolving fitness distributions are to be assessed. 


By relaxing strong assumptions on the frequency and effect of new mutations, our results go well beyond fitness-wave theory. As we saw with our numerical experiments, a generic evolutionary trajectory involves a phase of negative selection during which the effect of mutations on the fitness distribution is negligible. A consequence of this is that the low-fitness tail of the distribution becomes fatter than Gaussian. This means that \textit{there are in general many more low-fitness individuals in a population than predicted by fitness wave theory}. Reciprocally, our results suggest that measuring the skewness of the fitness distribution is a test of positive vs. negative selection: if this measured skewness is significantly negative, we can say that the population is undergoing negative selection. The $\beta$ parameter in Theorem 2 is a quantitative measure of such ``selection negativity". 

 Our approach also highlights the difference between universal and system-dependent features of evolving fitness distributions. Like temperature in Boltzmann's H-theorem or the mean and variance in the central limit theorem, we find that the mean, the variance and (to a lesser extent\footnote{The skewness of evolving fitness distributions must belong to the interval $[-2,0]$.}) the skewness---the first three cumulants---of evolving fitness distributions depend on the system under study, its underlying fitness landscape, etc.; all higher cumulants, on the other hand, are completely fixed once the latter are. This observation resolves a long-standing debate regarding the content and interpretation of Fisher's famous ``fundamental theorem of natural selection". It also constitute a definite statistical prediction of Darwin's theory of evolution through natural selection; empirical tests of this prediction with published microbiological fitness data are in progress and will be reported elsewhere. 


\acknowledgements

We thank M. Harper, A. Riello and S. A. Frank for useful critical comments on an early version of this manuscript. Research at the Perimeter Institute is supported in part by the Government of Canada through Industry Canada and by the Province of Ontario through the Ministry of Research and Innovation.

\bibliographystyle{utcaps}

\providecommand{\href}[2]{#2}\begingroup\raggedright\endgroup

\newpage
\appendix

	\section{Evolution as transport}\label{appendixevo}
The replicator-mutator equation in fitness space reads 
\begin{equation}\label{flow}
	\dot{p}_t(x)=(x-\mu_t)p_t(x)+U\int d\Delta\, m(\Delta)\,[p_t(x-\Delta)-p_t(x)]
\end{equation}
where $p_t(x)$ is the distribution of Malthusian fitness $x$, $\mu_t$  the mean fitness at time $t$, $U$ is the mutation rate, $m(\Delta)$ a fixed distribution of fitness effects $\Delta$ of new mutations and dot means $\partial/\partial t$. In the limit where these mutations are infinitely frequent with infinitely small effects, this equation can be reduced to a reaction-diffusion equation, as in Refs. \cite{Tsimring:1996cr,Rouzine:2003en}. This (usually unrealistic) approximation is unnecessary, however, as \eqref{flow} can be solved exactly in terms of generating functions.

Let
\begin{equation}
	\chi_t(s)=\int dx\, e^{sx}p_t(x)
\end{equation} 
be the moment-generating-function (MGF) of $p_t(x)$, such that $\mu_t=\chi_t'(0)$, and let $\psi_t(x)=\ln \chi_t(s)$ be its cumulant-generating-function (CGF). Denote $\chi_m(s)$ and $\psi_m(s)$ the corresponding quantities for the the DFE  $m(\delta)$. From \eqref{flow} we have
\begin{equation}
	\dot{\chi}_t(s)=\chi'_t(s)-\chi'_t(0)\chi_t(s)+U(\chi_m(s)-1)\chi_t(s),
\end{equation}
where prime means $\partial/\partial s$, hence
\begin{equation}\label{transport}
	\dot{\psi}_t(s)=\psi'_t(s)-\psi'_t(0)+U(\chi_m(s)-1).
\end{equation}
This transformed version of the replicator-mutator equation has two remarkable features: it is \textit{linear} and, except for the second term on the right-hand side, it has the structure of a \textit{transport equation}.\footnote{A transport equation in one-dimension is a first-order PDE of the form $\dot{f}+vf'=g$, where $v$ and $g$ are two functions called drift and source respectively.} This observation suggests that its general solution can be written in terms of the characteristics $s+t=\textrm{constant}$, and indeed we can check that 
\begin{equation}\label{sol}
	\psi_t(s)=\psi_0(s+t)-\psi_0(t)+U\int_0^tdu\, \big(\chi_m(s+u)-\chi_m(u)\big)
\end{equation}
is the general solution of \eqref{transport}. The interpretation which results is compelling: the evolution of fitness distributions through selection and mutations is equivalent to the transport of their CGF towards progressively lower values of $s$, with sources given by the mutational MGF. This is illustrated in Fig. \ref{transportfig}. 

The solution \eqref{sol} can be used to write an explicit formula for the evolved fitness distribution $p_t(s)$ using an inverse Laplace transform, namely 

\begin{equation}
	p_t(s)=\frac{1}{2i\pi}\int_{c-i\infty}^{c+i\infty} ds\, \frac{\chi_0(s+t)}{\chi_0(t)}\, \exp\left(U\int_0^tdu\, \big(\chi_m(s+u)-\chi_m(u)\big)-st\right).
\end{equation}

More usefully for our purposes, it also gives its cumulants $\kappa_t^{(p)}\equiv \psi_t^{(p)}(0)$ as
\begin{equation}\label{cum}
	\kappa_t^{(p)}=\psi_0^{(p)}(t)+U\big(\chi_m^{(p-1)}(t)-\chi_m^{(p-1)}(0)\big)\quad\textrm{for}\ \ p\geq 1.
\end{equation}
Formula \eqref{cum}, and its decomposition into pre-existing variation (first term on the RHS) and new mutations (second term on the RHS) is the basis for all our limit theorems.
	\section{Discrete time}

All steps above can be repeated in the discrete-time formulation the replicator-mutator dynamics. In this case we use an integer generation label $n$ instead of time $t$, and assume that an individual with fitness $x$ has $w=e^x$ offsprings per generation. (The quantity $w$ is the Wrightian fitness.) Then we can write

\begin{equation}
	p_{n+1}(x)=\frac{\int d\Delta\, M(\Delta)\, e^{x-\Delta}p_n(x)}{\int dx'e^{x'}p_n(x')}.
\end{equation}
with $M(\delta)\equiv (1-U)\delta(\Delta)+Um(\Delta)$. In terms of CGFs this reads
\begin{equation}
	\psi_{n+1}(s)=\psi_{n}(s+1)-\psi_n(1)+\ln\left[(1-U)+U\chi_m(s)\right]
\end{equation}
which, as already noted by Eshel \cite{Eshel:1971ur}, can be solved by a straightforward recursion:
\begin{equation}
	\psi_n(s)=\psi_0(s+n)-\psi_0(n)+\sum_{j=0}^{n-1}\ln\left[(1-U)+U\chi_m(s+j)\right].
\end{equation}
As before the cumulants of the fitness distribution follow immediately.

\section{Proofs}\label{proofs}

The proofs of Theorems 1-4 all follow the same pattern:

\begin{enumerate}
	\item Consider the \textit{standardized} (``renormalized", ``comoving") fitness distribution 
	\begin{equation}
		g_t(x)\equiv \sigma_t\, p_t(\sigma_t x+\mu_t),
	\end{equation}
	in which $\mu_t$ and $\sigma_t$ are the mean and standard deviation of $p_t(x)$. The cumulants $K_t^{(p)}$ of $g_t(x)$ are given by 
	\begin{equation}\label{standardizedcum}
		K_t^{(p)}\equiv \frac{\kappa_t^{(p)}}{\sigma_t^p}=\frac{\psi_t^{(p)}(0)}{\psi_t''(0)^{p/2}}.
	\end{equation}
	\item Use asymptotic estimates of generating functions for $s\to\infty$ to obtain the limit of \eqref{standardizedcum}	as $t\to\infty$.
	\item Identify the limit distribution $\lim_{t\to\infty} g_t(x)$ from its cumulants $\lim_{t\to\infty} K_t^{(p)}$.
	
\end{enumerate}
\subsection{Theorem 1}
\paragraph*{Asymptotic normality.}
The Kasahara Tauberian theorem \cite{Kasahara:1978wd} states that, if $p_0(x)$ is a distribution with 
\begin{equation}
	-\ln\int_x^\infty dy\,p_0(y)\underset{x\to\infty}{\sim} x^\alpha,
\end{equation}
for some $\alpha>1$, then its associated CGF $\psi_0(s)$ satisfies
\begin{equation}\label{asympCGF}
	\psi_0(s)\underset{s\to\infty}{\sim} s^{\overline{\alpha}}
\end{equation}
where $\overline{\alpha}$ is the exponent conjugate to $\alpha$, i.e. $1/\alpha+1/\overline{\alpha}=1$; see also \cite[Chap. 4.12]{BINGHAM:1989ca}. For fitness distributions in the regime of selection of pre-existing variation, we have $\psi_t(s)=\psi_0(s+t)-\psi_0(s)$, hence from \eqref{standardizedcum},
\begin{equation}\label{conv}
	 K_t^{(p)}\underset{t\to\infty}{\sim}t^{\overline{\alpha}(1-p/2)}.
\end{equation}
Since $\overline{\alpha}>1$, the exponent on the right-hand side is negative for all $p\geq 3$, hence 
\begin{equation}
	K_t^{(1)}=0,\quad K_t^{(2)}=1\quad \textrm{and}\quad\lim_{t\to\infty} K_t^{(p)}=0\ \textrm{for}\ p\geq3. 
\end{equation}
This implies that $g_t(x)$ converges (weakly hence uniformly\footnote{Weak convergence to a continuous distribution implies uniform convergence.}) to the standard normal distribution. Moreover we have
\begin{equation}
	\mu_t=\psi_0'(t)\underset{t\to\infty}{\sim} t^{\overline{\alpha}-1}\quad\textrm{and}\quad \sigma_t^2=\psi_0''(t)\underset{t\to\infty}{\sim} t^{\overline{\alpha}-2},
\end{equation}
as announced in Theorem 1.
\paragraph*{Rate of convergence.}
We close by showing that the rate of uniform convergence to the normal is $\mathcal{O}(t^{-\overline{\alpha}/2})$, following the steps in the proof of the Berry-Esseen theorem in \cite[Chap III-6]{Shiryaev:2013kv}. For this purpose we introduce the cumulative distribution functions 
\begin{equation}
	G_t(x)\equiv\int_{-\infty}^xdx'\, g_t(x')\quad\textrm{and}\quad\mathcal{N}(x)\equiv\int_{-\infty}^xdx'\, \frac{e^{-x'^2/2}}{\sqrt{2\pi}}.
\end{equation}
We start from Esseen's inequality, according to which for any $T>0$, 
	\begin{equation}\label{esseen}
		\sup_{x\geq0} \vert G_t(x)-\mathcal{N}(x)\vert \leq \frac{2}{\pi}\int_0^T\frac{\vert f_t(u)-e^{-u^2/2}\vert}{u}\,du+\frac{24}{\pi T\sqrt{2\pi}},
	\end{equation}
where $f_t(u)\equiv\exp \sum_{p=2}^{\infty}K_t^{(p)}(iu)^p/p!$ is the characteristic function of the standardized fitness distribution. Next, we use $\vert e^{z}-1\vert\leq \vert z\vert\, e^{\vert z\vert}$ to note that 
\begin{equation}\label{useful}
	\vert f_t(u)-e^{-u^2/2}\vert\leq e^{-u^2/2} \epsilon_t(u)\,e^{\epsilon_t(u)}
\end{equation}
with 
\begin{equation}
	\epsilon_t(u)\equiv\Big\vert \sum_{p=3}^{\infty}K_t^{(p)}\frac{(iu)^p}{p!}\Big\vert\leq u^3\,\vert K_t^{(3)}\vert\, \sum_{q=0}^{\infty}\Big\vert\frac{K_t^{(q+3)}}{K_t^{(3)}}\Big\vert \frac{T^q}{(q+3)!}\quad\textrm{for}\ u\in[0,T].
\end{equation}
 From \eqref{conv}, we have $K_t^{(3)}=\mathcal{O}(t^{-\overline{\alpha}/2})$ and 
\begin{equation}
	\Big\vert\frac{K_t^{(q+3)}}{K_t^{(3)}}\Big\vert=\mathcal{O}(t^{-q\overline{\alpha}/2})\prod_{k=0}^{q+2}(\overline{\alpha}-q).
\end{equation}
The series 
\begin{equation}
	\sum_{q=0}^{\infty}\frac{1}{(q+3)!}\prod_{k=0}^{q+2}(\overline{\alpha}-q)
\end{equation}
being summable, we can take $T=C_1n^{\overline{\alpha}/2}$ to obtain 
\begin{equation}
	\epsilon_t(u)=u^3\mathcal{O}(t^{-\overline{\alpha}/2}). 
\end{equation}
Let $C_2$ be a constant such that $\epsilon_t(u)\leq C_2 u^3 t^{-\overline{\alpha}/2}$. We have 
\begin{equation}
	-\frac{u^2}{2}+\epsilon_t(u)\leq -\frac{u^2}{4}+u^2\left(C_2 T n^{-\overline{\alpha}/2}-\frac{1}{4}\right)\leq -\frac{u^2}{4}+\left(C_1C_2-\frac{1}{4}\right),
\end{equation}
hence choosing $C_1=1/4C_2$ gives 
\begin{equation}
	-\frac{u^2}{2}+\epsilon_t(u)\leq -\frac{u^2}{4}. 
\end{equation}
Combining \eqref{useful} with \eqref{esseen}, we have obtained
\begin{equation}
	\sup_{x\geq0} \vert G_t(x)-\mathcal{N}(x)\vert \leq\mathcal{O}(t^{-\overline{\alpha}/2})\int_0^{C_1n^{\overline{\alpha}/2}}u^2 e^{-u^2/4}\, du+\mathcal{O}(t^{-\overline{\alpha}/2}).
\end{equation} 
The integral is bounded from above by $\int_0^{\infty}t^2 e^{-t^2/4}\, dt=2\sqrt{\pi}$ hence is $\mathcal{O}(1)$, and the conclusion follows.

\subsection{Theorem 2}

In the previous section we excluded the case $\overline{\alpha}=1$ in \eqref{asympCGF}. By the Paley-Wiener theorem, this case corresponds to fitness distributions with compact support (that is, with finite right endpoint $x_+$). If $p(x_+)\neq 0$, it is easy to see using Laplace's method that
\begin{equation}
		\chi(s)=\int_0^{x_+} e^{sx}p(x)dx\underset{s\to\infty}{\sim}  \frac{e^{sx_+}}{s}.
	\end{equation}
In particular, 
\begin{equation}
\psi(s)=\ln\chi(s)\underset{s\to\infty}{\sim}  x_+s-\ln s.
\end{equation}
It follows that the standardized cumulants in \eqref{standardizedcum} satisfy
\begin{equation}
	K_n^{(p)}\underset{t\to\infty}{\sim}  \frac{-\ln^{(p)}(t)}{\ln''(t)^{p/2}}=(-1)^p(p-1)!. 
\end{equation}
This sequence of moments characterizes the reversed exponential distribution $e^{-1+x}$, to which $g_t$ therefore converges (weakly and uniformly) as $t\to\infty$. (This can be seen by resumming the cumulant series, which gives the moment-generating function $e^{s}/(1+s)$.) 

This computation can be generalized to the case where $p(x_+)=0$. Suppose there exists $\beta>0$ such that\footnote{We conjecture that it is sufficient that $F(x)$ is regularly varying at $x=x_+$ with index $\beta+1$, but do not know how to prove the theorem in this case.}
\begin{equation}
	p_0(x)\underset{x\to x_+}{\sim}  (x_+-x)^{\beta}
\end{equation}
Then a refined version of Laplace's method \cite[Chap 5-2]{Bleistein:1975vr} gives 
\begin{equation}
	\chi_0(s)=\int_0^{x_+} e^{sx}p(x)dx\underset{s\to\infty}{\sim} \frac{e^{sx_+}}{s^{1+\beta}}
\end{equation}
It follows that 
\begin{equation}\label{asympfinite}
\psi_t(s)\underset{s\to\infty}{\sim} x_+s-(1+\beta)\ln s	
\end{equation}
hence for the standardized cumulants 
\begin{eqnarray}
	K_t^{(p)}\underset{t\to\infty}{\sim} (1+\beta)^{1-p/2}\frac{-\ln^{(p)}(t)}{\ln''(t)^{p/2}}=(1+\beta)^{1-p/2} (-1)^p\,(p-1)!. 
\end{eqnarray}
Resumming the cumulant series gives the CGF
\begin{eqnarray}
	\sum_{p=0}^\infty K_t^{(p)}\frac{s^p}{p!}&\underset{t\to\infty}{\sim}&(1+\beta)\sum_{p=2}^\infty \frac{(-1)^p}{p}\,(1+\beta)^{-p/2}s^p\\&=&(1+\beta)\Big[s(1+\beta)^{-1/2}-\ln\Big(1+s(1+\beta)^{-1/2}\Big)\Big].
\end{eqnarray}
We can obtain the associated density function $\lim_{t\to\infty}g_{t}(x)$ by means of an inverse Laplace transform, which gives
\begin{equation}\label{newdistribution}
	\lim_{t\to\infty}g_{t}(x)=\frac{(1+\beta)^{(1+\beta)/2}}{\Gamma(1+\beta)}\,e^{-(1+\beta)^{1/2}[(1+\beta)^{1/2}-x]}\Big[(1+\beta)^{1/2}-x\Big]^\beta\quad\textrm{for}\ x<(1+\beta)^{1/2}.
\end{equation}
This, together with the observation that 
\begin{equation}
	\mu_t=\psi_0'(t)\underset{t\to\infty}{\sim} x_+\quad\textrm{and}\quad \sigma_t^2=\psi_0''(t)\underset{t\to\infty}{\sim} t^{-2},
\end{equation}
concludes the proof of Theorem 2. 

\subsection{Theorem 3}

In the regime of positive selection of new mutations, \eqref{cum} implies for the standardized cumulants 
\begin{equation}\label{cummut}
	K_t^{(p)}=U^{1-p/2}\, \frac{\chi_m^{(p-1)}(t)-\chi_m^{(p-1)}(0)}{[\chi_m'(t)-\chi_m'(0)]^{p/2}}.
\end{equation}
We assume for simplicity that the DFE has a bounded support with finite right end-point $\Delta_+>0$. Using the same Laplace estimate as above, we have
\begin{equation}
	\chi_m(s)\underset{s\to\infty}{\sim}\frac{e^{s\Delta_+}}{s^{1+\beta}}.
\end{equation}
It follows that 
\begin{equation}
	K_t^{(p)}\underset{t\to\infty}{\sim}\left(U\,\frac{e^{\Delta_+s}}{s^{1+\beta}}\right)^{1-p/2}
	\end{equation}
	which goes to zero for all $p\geq3$. Thus $g_t$ converges again to a normal distribution. 
	
	\subsection{Theorem 4}
	
When all mutations are deleterious, $\Delta_+<0$, the first in  \eqref{cummut} goes to zero as $t\to\infty$, hence 
\begin{equation}
	\lim_{t\to\infty}K_t^{(p)}=-U^{1-p/2}\,\frac{\chi_m^{(p-1)}(0)}{[-\chi_m'(0)]^{p/2}}
\end{equation} 
as claimed in Thm. 4. 
	\section{Wright-Fisher simulations}\label{WFappendix}
		
	The Wright-Fisher process is a well-known stochastic model of asexual evolution through selection and mutations in finite populations. It is based on non-overlapping generations, whose composition is determined probabilistically by $(i)$ the composition of the earlier generation, and $(ii)$ a fixed mutation rate $U$.
	
	In this work we set up a Wright-Fisher process in the following way. First, we consider a discretized fitness space consisting of $F$ bins $i$ with Wrightian fitness (number of offsprings)  $w(i)=w_{\textrm{min}} +(w_{\textrm{max}}-w_{\textrm{min}})i/F$. The number of individuals with fitness $f_i$ at generation $n$ is denoted $N_n(i)$, and the total population $N=\sum_iN_n(i)$ is fixed. Finally we assume that a distribution of fitness effects has been given, in the form of fixed transition matrix $M(j,i)$ giving the probability of a mutation changing the fitness from $w(j)$ to $w(i)$. 
	
	The composition of generation $n$ then determines the probability that a randomly chosen individual at generation $n+1$ is the fitness bin $i$ as
	
	\begin{equation}
			p_{n+1}(i)=\frac{(1-U)w(i) N_{n}(i)+U\sum_j M(j,i)w(j)N_{n}(j)}{\sum_i[ (1-U)w(i) N_{n}(i)+U\sum_j M(j,i)w(j)N_{n}(j)]}.
	\end{equation}
	In words, this individual is either (with probability $1-U$) the non-mutated offspring of a parent in the same bin $i$, which comes with a weight $w(i) N_{n}(i)$, or (with probability $U$) the mutated offspring of a parent in one of the other bins $j$, which comes with a weight $M(j,i)w(j)N_n(j)$. The probabilities $p_{n+1}	(i)$ then determine the probability of the number distribution $\{N_{n+1}\}$ according to the multinomial expression: 
	\begin{equation}
		\textrm{Prob}(\{N_{n+1}\})=\frac{N!}{\prod_i N_{n+1}(i)}\prod_i p_{n+1}(i)^{N_{n+1}(i)}.
	\end{equation}

	In the main text we present the results of our simulations with $F=500$, $U=10^{-2}$, total population $N=10^5$ and a transition matrix of the form
	\begin{equation}
		M(j,i)=h\left(\frac{w(i)}{w(j)}-1\right),
	\end{equation}
	where $h$ is a Laplace distribution $h(z)=e^{-\vert x-\mu_m\vert/\sigma_m}/2\sigma_m$.

	\section{Genetic algorithm}\label{GAappendix}
	
We ran a genetic algorithm (GA) to test the applicability of our Theorems in a setting where the notion of ``distribution of fitness effects of new mutations" is not \textit{a priori} well-defined. The GA performs an iterative search for the solution of the linear optimization problem
\begin{equation}
	\max\{c\cdot y;\, y\in\{1,\cdots,Q\}^L\ \textrm{and}\ b\cdot y\leq d\}
\end{equation}
where $L,Q\in\mathbb{N}$ and $d>0$ are fixed numbers, $b$ and $c$ are two randomly chosen vectors in $[0,1]^L$, and $\cdot$ denotes the dot product of vectors. 

Our GA proceeds by maintaining a population of $N$ strings $y$ (``genes") of $L$ integers $y_i\in\{1,\cdots,Q\}$ (``bases"). Each genome is assigned the Wrightian fitness
\begin{equation}
	w(y)=\exp\left[\frac{\min(c\cdot y,\,[c\cdot y-(b\cdot y-d)]_+)}{a}\right]
	\end{equation}
where $[\,\cdot\,]_+\equiv \max(\,\cdot\,,0)$ and $a$ is fixed positive number ($1/a$ is the ``selective pressure"). The exponential function ensures that a random population of genes does \textit{not} has a fitness distribution resembling the attractors in Theorems 1-3.     

\begin{figure}[t]
	\includegraphics[scale=.8]{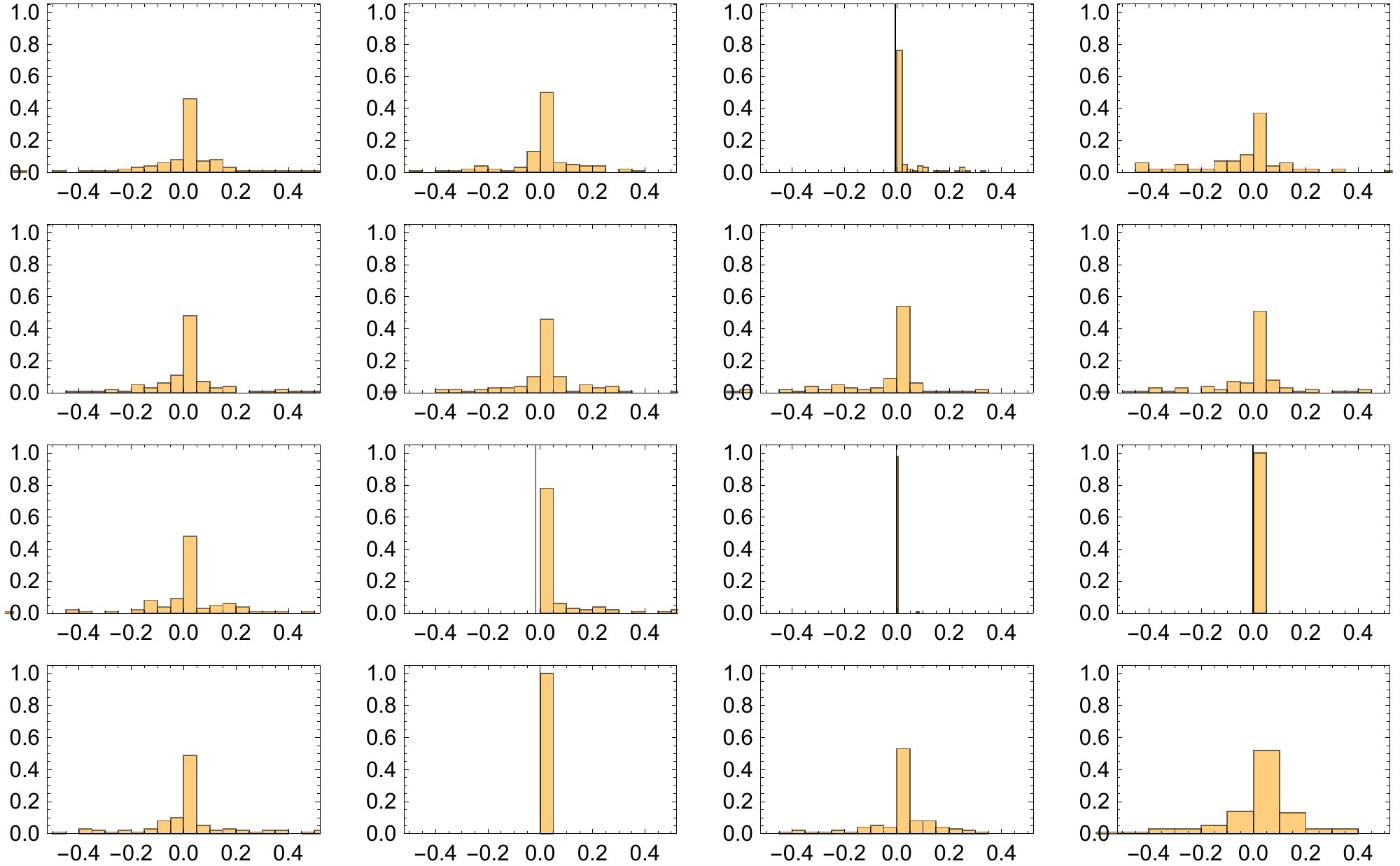}
	\caption{The GA does not have a single, well-defined DFE. In this figure each plot represents a histogram of the fitness effects $\Delta$ found in the mutational neighborhood of a randomly chosen gene. In some cases all mutations are almost neutral; in other cases they are mostly deleterious; in some rare cases all mutations are beneficial.}
	\label{gadfe}
\end{figure}

Then at each new generation the following two steps are taken:
\begin{enumerate}
	\item \textit{Selection.} A new population is generated by choosing $N$ genes $y$ from the previous populations weighted by their Wrightian fitness $w(y)$. 
	\item \textit{Mutation.} With probability $u<1$, each base of each gene is mutated into a randomly chosen new base in $\{1,\cdots,Q\}$.
\end{enumerate}
	At each generation the distribution $p_n(x)$ of Malthusian fitness $x(y)=\ln w(y)$ is measured and its cumulants $\kappa_n^{(p)}$ extracted. We also tested whether the GA has a single, well-defined DFE--- it does not, see Fig. \ref{gadfe}.

\section{Classical distributions}\label{defs}

For the reader's convenience we collect here the definitions of the classical probability distributions mentioned in the main text, in standard form. The following special functions are used:

\begin{itemize}
	\item The Gamma function, $$\Gamma(s)\equiv\int_0^\infty t^{s-1}e^{-s}\,dt$$
	\item The (lower) incomplete Gamma function, $$\gamma(s,x)\equiv\int_0^x t^{s-1}e^{-s}\,dt$$
	\item The Beta function, $$B(s,r)\equiv\int_0^1t^{s-1}(1-t)^{r-1}\,dt$$
	\item The incomplete Beta function, $$B(x;s,r)\equiv\int_0^xt^{s-1}(1-t)^{r-1}\,dt$$
	\item The Owen T function, $$T(h;a)\equiv\frac{1}{2\pi}\int_0^a\frac{e^{-h^2(1+t^2)/2}}{1+t^2}	 \,dt$$

\end{itemize}

\setlength{\tabcolsep}{.6em}

\begin{sidewaystable}
\begin{center}
{\renewcommand{\arraystretch}{1.3}	
{\large
\begin{tabular}{|c|c|c|c|c|c|}
   \hline
   Name  & Cumulative density function & Probability density function & Support & Shape parameters\\
   \hline

   Normal   & $\frac{1}{2}\left[1+\textrm{erf}(x/\sqrt{2})\right]$ & $\frac{1}{\sqrt{2\pi}}e^{-x^2/2}$ & $x\in\mathbb{R}$ & \\
   
      Skew-normal  & $\frac{1}{2}\left[1+\textrm{erf}(x/\sqrt{2})\right]-2T(x,\alpha)$ & $\frac{1}{\sqrt{2\pi}}e^{-x^2/2} \left[1+\textrm{erf}(\alpha x/\sqrt{2})\right]$ & $x>0$ & \\

   Log-normal  & $\frac{1}{2}\left[1+\textrm{erf}(\ln x/\sqrt{2})\right]$ & $\frac{1}{x\sqrt{2\pi}}e^{-(\ln x)^2/2}$ & $x>0$ & \\
%
%
        Gamma &  $\frac{1}{\Gamma(\alpha)}\,\gamma(\alpha,x)$ & $\frac{1}{\Gamma(\alpha)}\,x^{\alpha-1}e^{-x}$ & $x>0$ & $\alpha>0$\\
        Beta & $\frac{B(x;\alpha,\beta)}{B(\alpha,\beta)}$ & $\frac{x^{\alpha-1}(1-x)^{\beta-1}}{B(\alpha,\beta)}$ & $x\in(0,1)$ & $\alpha,\beta>0$ \\
        Weibull &  $1-e^{-x^\alpha}$ & $\alpha x^{\alpha-1}e^{-x^\alpha}$ & $x>0$ & $\alpha>0$\\
%
%
\hline
   \end{tabular}
   
}
}
\end{center}
\end{sidewaystable}

\end{document}